\documentclass[apj]{emulateapj}

\slugcomment{Accepted by The Astrophysical Journal on August 29, 2011}

\shorttitle{AGNs in zCOSMOS galaxy pairs}
\shortauthors{Silverman et al.}

\begin{document}


\title{The impact of galaxy interactions on AGN activity in zCOSMOS}


\author{J.~D.~Silverman \altaffilmark{1,2}, P.~ Kampczyk\altaffilmark{2}, K.~Jahnke\altaffilmark{3}, R.~Andrae\altaffilmark{3}, S.~J.~Lilly\altaffilmark{2}, M.~Elvis\altaffilmark{4}, F.~Civano\altaffilmark{4},  V. Mainieri\altaffilmark{5}, C. Vignali \altaffilmark{6}, G. Zamorani\altaffilmark{7}, P.~Nair\altaffilmark{7}, O.~ Le F\`{e}vre\altaffilmark{8}, L.~de~Ravel\altaffilmark{9}, S.~Bardelli\altaffilmark{7}, A.~Bongiorno\altaffilmark{10}, M.~Bolzonella\altaffilmark{7}, A.~Cappi\altaffilmark{7}, K.~Caputi\altaffilmark{11}, C. M.~Carollo\altaffilmark{2}, T.~Contini\altaffilmark{12,13}, G.~ Coppa\altaffilmark{7}, O.~ Cucciati\altaffilmark{8}, S.~ de la Torre\altaffilmark{8}, P.~ Franzetti\altaffilmark{14}, B.~ Garilli\altaffilmark{14}, C. Halliday\altaffilmark{15}, G.~Hasinger\altaffilmark{16}, A.~ Iovino\altaffilmark{17}, C. ~Knobel\altaffilmark{2}, A.~Koekemoer\altaffilmark{18}, K.~ Kova\v{c}\altaffilmark{2}, F.~Lamareille\altaffilmark{8}, J.~-F. Le Borgne\altaffilmark{12,13}, V.~ Le Brun\altaffilmark{8}, C.~Maier\altaffilmark{2}, M.~ Mignoli\altaffilmark{7}, R.~ Pello\altaffilmark{12,13}, E.~ Perez Montero\altaffilmark{12,13}, E.~ Ricciardelli\altaffilmark{19}, Y.~ Peng\altaffilmark{2}, M.~ Scodeggio\altaffilmark{17}, M.~ Tanaka\altaffilmark{1}, L.~ Tasca\altaffilmark{17}, L.~ Tresse\altaffilmark{8}, D.~ Vergani\altaffilmark{7}, E.~ Zucca\altaffilmark{7}, M.~Brusa\altaffilmark{10}, N.~Cappelluti \altaffilmark{7}, A.~Comastri\altaffilmark{7}, A.~Finoguenov\altaffilmark{10}, H. Fu\altaffilmark{20}, R.~Gilli\altaffilmark{7}, H.~Hao\altaffilmark{3}, L. C. Ho\altaffilmark{21}, M.~Salvato\altaffilmark{14}
}


\altaffiltext{1}{Institute for the Physics and Mathematics of the Universe (IPMU), University of Tokyo, Kashiwanoha 5-1-5, Kashiwa-shi, Chiba 277-8568, Japan}
\altaffiltext{2}{Institute of Astronomy, ETH Z\"urich, CH-8093, Z\"urich, Switzerland.}
\altaffiltext{3}{Max-Planck-Institut f\"ur Astronomie, K\"onigstuhl 17, D-69117 Heidelberg, Germany}
\altaffiltext{4}{Harvard-Smithsonian Center for Astrophysics, 60 Garden Street, Cambridge, MA, 02138}
\altaffiltext{5}{European Southern Observatory, Karl-Schwarzschild-Strasse 2, Garching, D-85748, Germany}
\altaffiltext{6}{Dipartimento di Astronomia, Universit\'a di Bologna, via Ranzani 1, I-40127, Bologna, Italy}
\altaffiltext{7}{INAF Osservatorio Astronomico di Bologna, via Ranzani 1, I-40127, Bologna, Italy}
\altaffiltext{8}{Laboratoire d'Astrophysique de Marseille, Marseille, France}
\altaffiltext{9}{Institute for Astronomy, University of Edinburgh, Royal Observatory, Edinburgh, EH93HJ, UK}
\altaffiltext{10}{Max-Planck-Institut f\"ur extraterrestrische Physik, D-84571 Garching, Germany}
\altaffiltext{11}{SUPA, Institute for Astronomy, The University of Edinburgh, Royal Observatory, Edinburgh EH9 3HJ, UK}
\altaffiltext{12}{IRAP, Universit de Toulouse, UPS-OMP, Toulouse, France}
\altaffiltext{13}{CNRS, IRAP, 14, avenue Edouard Belin, F-31400 Toulouse, France}
\altaffiltext{14}{INAF - IASF Milano, 20133 Milan, Italy}
\altaffiltext{15}{INAF Osservatorio Astronomico di Arcetri, 50125 Firenze, Italy}
\altaffiltext{16}{Max-Planck-Institut fuer Plasmaphysik, Boltzmannstrasse 2, 85748 Garching bei Muenchen, Germany}
\altaffiltext{17}{INAF Osservatorio Astronomico di Brera, Milan, Italy}
\altaffiltext{18}{Space Telescope Science Institute, 3700 San Martin Drive, Baltimore, MD, 21218, USA}
\altaffiltext{19}{Dipartimento di Astronomia, Universita di Padova, Padova, Italy}
\altaffiltext{20}{California Institute of Technology, Pasadena, CA, USA.}
\altaffiltext{21}{The Observatories of the Carnegie Institution for Science, 813 Santa Barbara Street, Pasadena, CA 91101, USA}


\begin{abstract}

Close encounters between galaxies are expected to be a viable mechanism, as predicted by numerical simulations, by which accretion onto supermassive black holes can be initiated.  To test this scenario, we construct a sample of 562 galaxies ($M_*>2.5\times10^{10}$ M$_{\sun}$) in kinematic pairs over the redshift range $0.25<z<1.05$ that are more likely to be interacting than a well-matched control sample of 2726 galaxies not identified as being in a pair, both from the zCOSMOS 20k spectroscopic catalog.  Galaxies that harbor an active galactic nucleus (AGN) are identified on the basis of their X-ray emission ($L_{0.5-10~{\rm keV}}>2\times10^{42}$ erg s$^{-1}$) detected by $Chandra$.  We find a higher fraction of AGN in galaxies in pairs relative to isolated galaxies of similar stellar mass.  Our result is primarily due to an enhancement of AGN activity, by a factor of 1.9 (observed) and 2.6 (intrinsic), for galaxies in pairs of projected separation less than 75 kpc and line-of-sight velocity offset less than 500 km s$^{-1}$.  This study demonstrates that close kinematic pairs are conducive environments for black hole growth either indicating a causal physical connection or an inherent relation, such as, to enhanced star formation.  In the Appendix, we describe a method to estimate the intrinsic fractions of galaxies (either in pairs or the field) hosting an AGN with confidence intervals, and an excess fraction in pairs.  We estimate that $17.8_{-7.4}^{+8.4}\%$ of all moderate-luminosity AGN activity takes place within galaxies undergoing early stages of interaction that leaves open the question as to what physical processes are responsible for fueling the remaining $\sim80\%$ that may include late-stage mergers.
 
\end{abstract}





\section{Introduction}

There has been a long-standing question in astrophysics, ever since quasars and AGN were firmly believed to be powered by accretion \citep{salpeter1964} onto supermassive black holes (SMBH);  what physical mechanism(s) is (are) responsible for the loss of angular momentum of gas that is initially rotationally-supported and that can then be available to fuel a central SMBH \citep[e.g.,][]{lynden-bell1971}?  There have been a number of proposed mechanisms that have been investigated, over the past few decades, such as disk instabilities, galaxy mergers, or supernova-driven winds, just to name a few.  More recently, galaxy mergers, including interactions between galaxies with larger separations, have been a leading contender since numerical simulations demonstrate that such events \citep{mihos1996} can generate mass inflow to the nuclear region thus potentially fueling both AGNs and central starbursts \citep[e.g., ][]{hopkins2008}.  

In support of the merger scenario, the by-products of major mergers of gas-rich galaxies, the ultra-luminous infrared galaxies (ULIRGs), have been known for decades to be conducive to black hole growth \citep{sanders1996}.  ULIRGS have been shown to have high nuclear concentrations of gas \citep{scoville1989} that likely provide an ample fuel reservoir for accretion onto a SMBH.  Indeed, the incidence of AGN in ULIRGs \citep[e.g.,][]{kartaltepe2010} rises substantially with IR luminosity to an unprecedented level ($>50\% $) for the brightest that reach bolometric luminosities equivalent to quasars.  

There has been much effort \citep[e.g.,][]{bahcall1997,canalizo2001} over the past decade to determine whether galaxy interactions and mergers play a role in fueling quasars by carrying out imaging campaigns of their host galaxies with HST to look for signs of morphological disturbances.  Such efforts are challenging because of the glaring light of a nuclear point source, the lack of adequate control samples, and the fact that the detection of such signs of interaction is highly redshift dependent \citep{kampczyk2007}.  Fortunately, new samples of AGNs have been constructed that appear to alleviate some of these difficulties.  In particular, X-ray surveys are identifying an obscured population \citep[e.g.,][]{barger2003,szokoly2004,silverman2010} that makes it easier to discern the morphology of their host.  Large-area surveys, with HST coverage and multi-wavelength support such as COSMOS \citep{sco07,ko2007}, are enabling the construction of adequate control samples.  To date, studies based on X-ray selected AGN have demonstrated that major mergers of galaxies are not likely to be the single dominant mechanism responsible for triggering AGN activity \citep{grogin2005,gabor2009, cisternas2011}, even though remarkable examples of a binary AGN have been found in the COSMOS field \citep{comerford2009,civano2010} and elsewhere \citep[e.g.,][]{green2010,shen2010,fu2010}.  One explanation may be that deep surveys mainly detect moderate-luminosity AGNs (i.e., ``Seyfert" population, $L_{X-ray}\sim10^{42-44}$ ergs s$^{-1}$) thus the fueling mechanisms may differ from the more luminous quasars \citep{hopkins2009}.

We present an alternative test of the hypothesis that galaxy interactions play a role in fueling moderate-luminosity AGNs.  We identify galaxies in the zCOSMOS spectroscopic survey that are in close physical proximity (i.e., galaxies in close kinematic pairs).  These are two galaxies within a given projected separation and line-of-sight velocity difference that likely indicates an interacting system that may merge at subsequent times.  This approach is independent of galaxy morphology thus unbiased in this respect.  We determine the fraction of galaxies in kinematic pairs that host an AGN based on X-ray emission detected by $Chandra$ and compare to a well-defined control sample.  It is important to recognize that the X-rays enable us to cleanly identify black hole accretion, less hampered by star formation that can severely impact optical emission-line studies, especially for galaxies in pairs.  We highlight that we are investigating the impact of interactions on scales between $\sim10-150$ kpc thus our study will provide constraints on early-stages of merger-driven models of black hole growth before a coalescence phase.  Conclusive evidence for such an enhancement of moderate-luminosity AGN activity, relative to the non-active galaxy population, has been elusive in SDSS samples at lower redshifts \citep[e.g.,][]{li2008,ellison2008}.  Throughout this work, we assume $H_0=70 $ km s$^{-1}$ Mpc$^{-1}$, $\Omega_{\Lambda}=0.75$, $\Omega_{\rm{M}}=0.25$, and AB magnitudes.

\section{zCOSMOS 20k galaxy redshift survey}

zCOSMOS \citep{lilly07,lilly09} is an ESO program to acquire spectroscopic redshifts for a large sample of galaxies in the COSMOS field \citep{sco07} with VIMOS on the VLT.  A 'bright' sample of 20,682 galaxies ($i<22.5$) is observed with a red grism to provide a wavelength coverage of $5500-9500$ ${\rm \AA}$ ideal for identifying $L_*$ galaxies up to $z \sim 1.2$.  We use this catalog to construct a well-defined sample of galaxies.  Specifically, we identify 15,807 galaxies with $i_{ACS} \leq 22.5$ and spectroscopic redshifts between $0.25<z<1.05$, each having a quality flag ($\ge1.5$) that amounts to a confidence of $\sim99\%$ in the redshift measurements for the overall sample.  We do not include any serendipitous objects that happen to fall within a slit.  Full details on data acquisition, reduction, and redshift measurements can be found in \citet{lilly07,lilly09}.

We further isolate 10,964 zCOSMOS galaxies that fall within  the $Chandra$ survey area \citep[see below for details, ][]{elvis2009} since we use the X-rays from this data set to identify those hosting an AGN.  Of these, there are 3481 with a stellar mass above $M_*>2.5\times10^{10}$ M$_{\sun}$ and $0.25<z<1.05$.  These limits are chosen to ensure a fairly complete representation of both blue and red galaxies over the given range in redshift \citep[see][]{si09b}.  Derived properties such as stellar mass and rest-frame color (Figure~\ref{color_mass}) are determined for each galaxy as detailed in \citet{bo10} and \citet{po10}.





\subsection{Kinematic pairs}

\begin{figure}
\epsscale{1.1}
\plotone{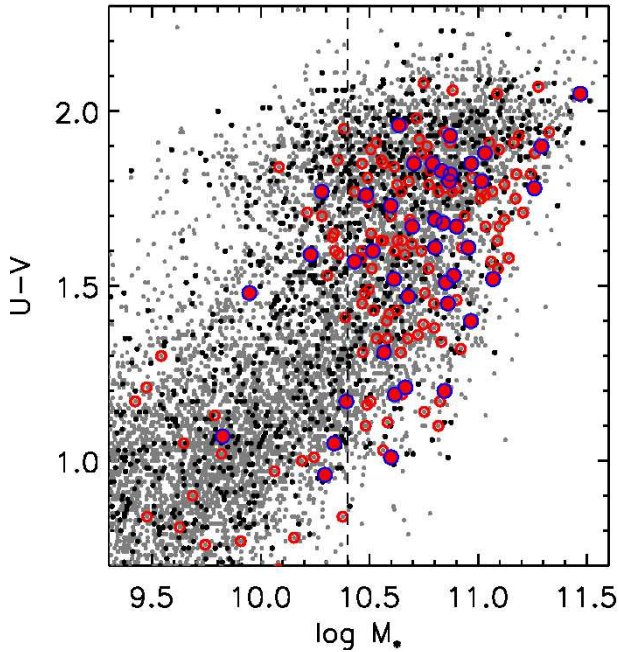}
\caption{Rest-frame color $U-V$ versus stellar mass for zCOSMOS galaxies with $0.25<z<1.05$ (small circles).  Galaxies in kinematic pairs are shown in black while all others are marked in grey.  The AGNs are further marked with colored circles (filled=kinematic pair; open=all others).  The vertical line marks our chosen mass limit.}
\label{color_mass}
\end{figure}

A major focus of the zCOSMOS survey is the study of galaxies in close kinematic pairs as a laboratory to determine the redshift evolution of the galaxy merger rate \citep{deravel2011}  and the physical properties of the galaxies themselves \citep{kampczyk2011}.  We use the zCOSMOS 20k catalog to identify those galaxies with a nearby neighbor.  Galaxies in kinematic pairs are specifically selected to have a  projected separation $dr<100$ kpc h$^{-1}$ (143 kpc for our chosen $h=0.7$) and a line-of-sight velocity difference $dV<500$ km s$^{-1}$ \citep{kampczyk2011}.  Based on repeat observations of 569 zCOSMOS galaxies, individual redshifts are accurate to $\sigma_z=0.00036\times(1+z)$ or $\sigma_z=110$ km s$^{-1}$ \citep[see][for details]{lilly09}, an accuracy well below our chosen velocity difference for identifying pairs.  Based on this selection, we identify 753 galaxies, out of the 3481 (selected by stellar mass and redshift) that are associated with a galaxy pair.  Of these, there are 330 with a mass ratio less than 3 to 1, while 562 have a mass ratio between 10 and 1.  In Figure~\ref{color_mass}, we show the stellar mass and rest-frame color of the galaxies in kinematic pairs.

\begin{figure}
\epsscale{1.1}
\plotone{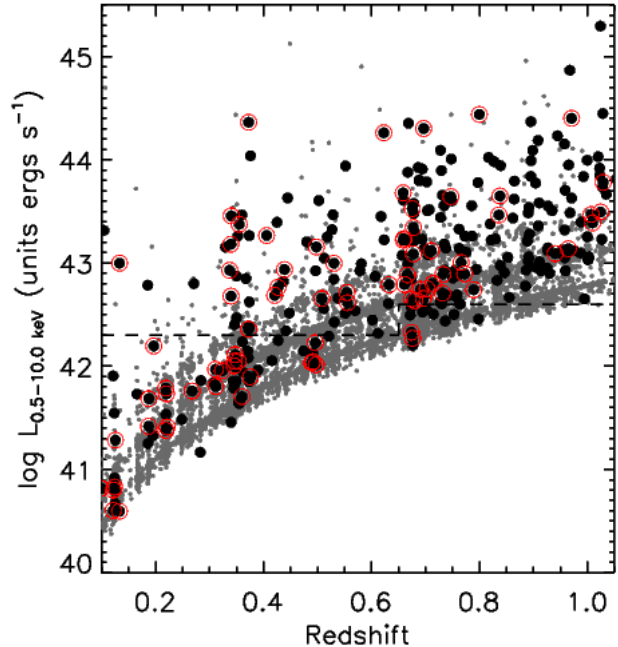} 
\caption{X-ray luminosity (0.5-10 keV) versus redshift for AGN (large symbols) with those in galaxy pairs further marked with an open, red circle.  The upper limit to the broad-band X-ray luminosity for each zCOSMOS galaxy is shown by the small grey dots.  The solid horizontal line is the minimum X-ray luminosity that we enforce to measure the fraction of galaxies hosting AGN in two redshift intervals.}
\label{lx_z}
\end{figure}

\section{$Chandra$ X-ray observations and AGN identification}
\label{chandra}

We match our zCOSMOS galaxy sample to the catalog of $Chandra$ X-ray sources \citep{elvis2009,puccetti2009} using a maximum-likelihood routine as detailed in \citet{civano2011}.  The $Chandra$ observations cover the central 0.9 sq. deg and result in the detection of 1761 point sources above a flux level of $5.7\times10^{-16}$ erg cm$^{-2}$ s$^{-1}$ in the 0.5-10 keV band.  The tiling scheme not only provides a fairly uniform sensitivity but an optimal PSF across the field.  We highlight that the resolving power of $Chandra$ is important with respect to our study since galaxy pairs have been found to reside in regions of heightened galaxy overdensities \citep{lin2010,deravel2011,kampczyk2011} where extended X-ray emission, if present, may hamper the detection of point sources.  The $Chandra$ observations over the C-COSMOS region are sensitive to AGNs with L$_{0.5-10~{\rm keV}}>2\times10^{42}$ ergs s$^{-1}$ out to $z\sim1$, a luminosity regime well-above that due to star formation thus indicative of a highly robust selection of AGNs.  

We identify 337 zCOSMOS galaxies with \hbox{X-ray} emission ($f_{0.5-10~{\rm keV}}>1\times10^{-15}$ erg cm$^{-2}$ s$^{-1}$) that primarily falls within the range $10^{42}\lesssim$L$_{0.5-10~{\rm keV}}\lesssim10^{44}$ erg s$^{-1}$ (Figure~\ref{lx_z}).  The X-ray luminosity of each AGN is determined from their observed broad-band (0.5-10.0 keV) flux assuming a power-law spectrum (photon index $\Gamma=2.0$) and no correction for intrinsic absorption.  In terms of Eddington ratio, we are sensitive to SMBHs having $10^{-3} \lesssim L_{\rm Bol}/L_{\rm Edd} \lesssim 10^{-1}$ given the stellar masses of their hosts, an assumption of their bulge-to-total mass ratio ($B/T \sim 0.5$), and the local $M_{BH}-M_{bulge}$ relation.  The majority of AGNs have optical emission dominated by their host galaxies and lack broad optical emission lines, both highlighting the importance of X-ray selection to account for the moderately-obscured population at these redshifts  \citep{mainieri2007,brusa2010}.

We compare the properties of our AGN hosts to the underlying galaxy population by selecting those with $10^{42.0}\lesssim$L$_{0.5-10~{\rm keV}}\lesssim10^{43.7}$ erg s$^{-1}$, a range that likely results in a purely AGN-dominated sample, in terms of X-ray emission, and avoids the inclusion of more optically-luminous AGNs that can impact the derived rest-frame colors and stellar mass estimates \citep[e.g.,][]{silverman2008}.  In Figure~\ref{color_mass}, we see that their host galaxies are massive ($M_*>2.5\times10^{10}$ M$_{\sun}$) and residual star formation may be present as evident by the slightly bluer colors ($<U-V>=1.6$) of AGN hosts as compared to non-active galaxies ($<U-V>=1.7$) of similar stellar mass \citep[e.g.,][]{si09b,schawinski2010,xue2010}.  This is in agreement with \citet{kampczyk2011} who find that galaxies in zCOSMOS kinematic pairs have elevated star formation rates, heightened post-starburst signatures and an increase in the numbers of galaxies with irregular morphologies.  

\section{Observed AGN fraction}
\label{obs_results}

\begin{figure}
\epsscale{1.1}
\plotone{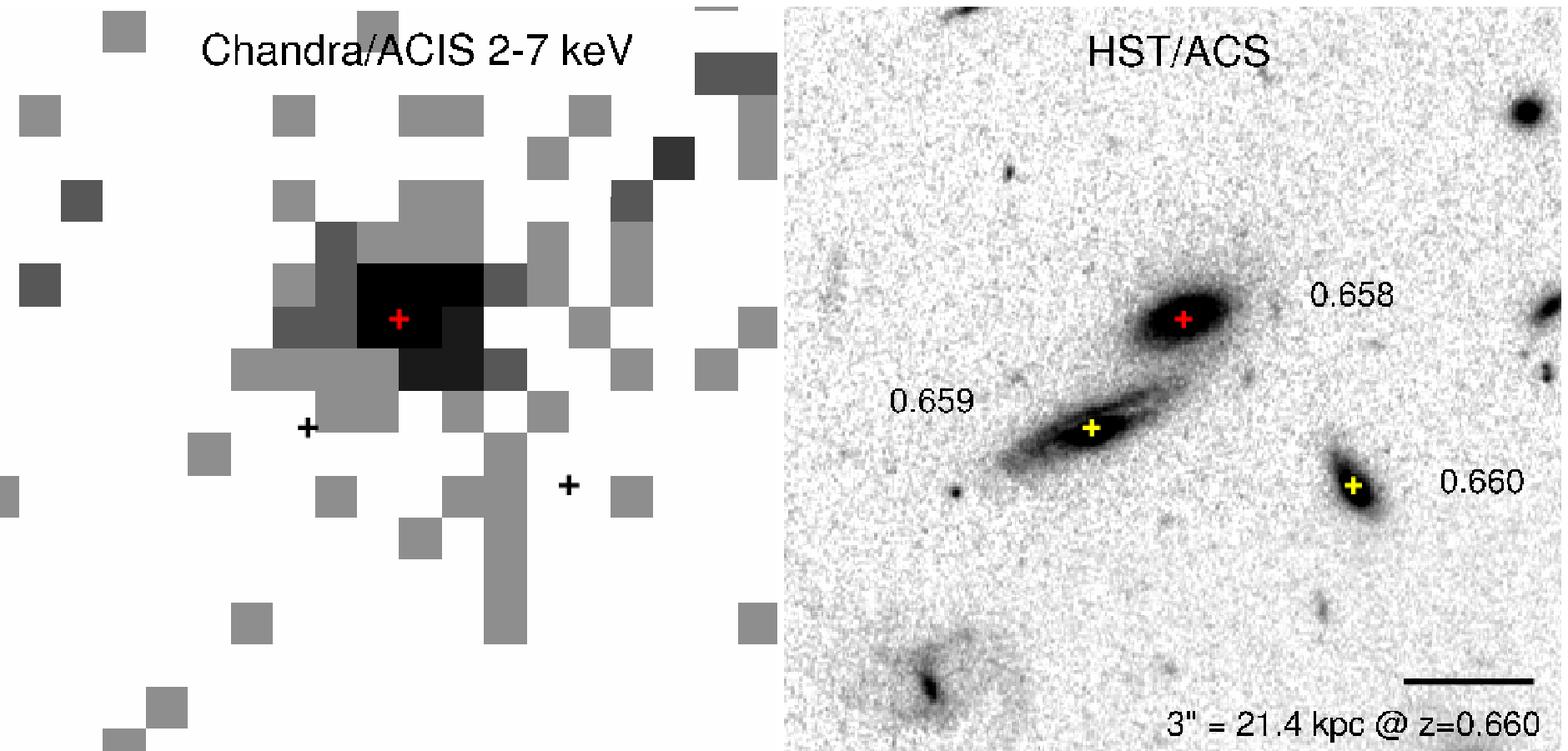}
\plotone{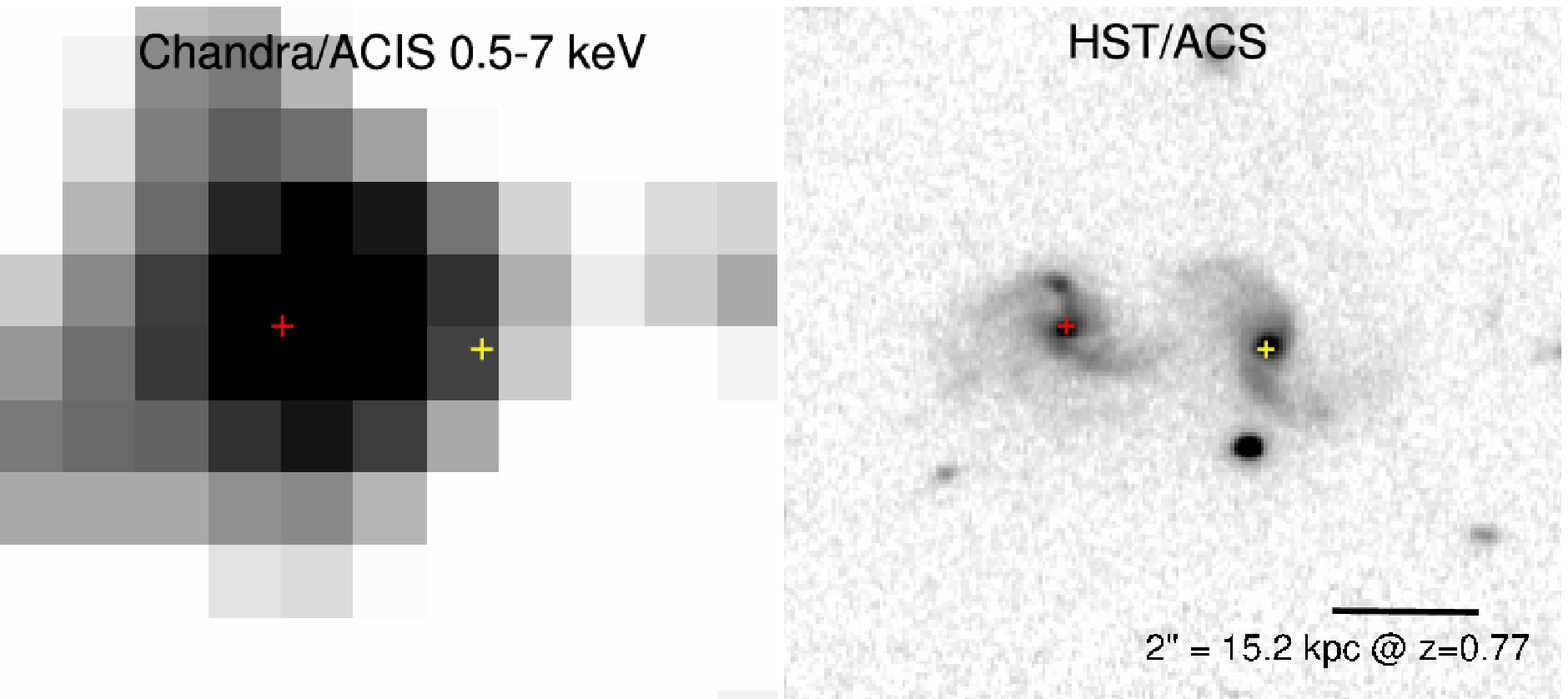}
\plotone{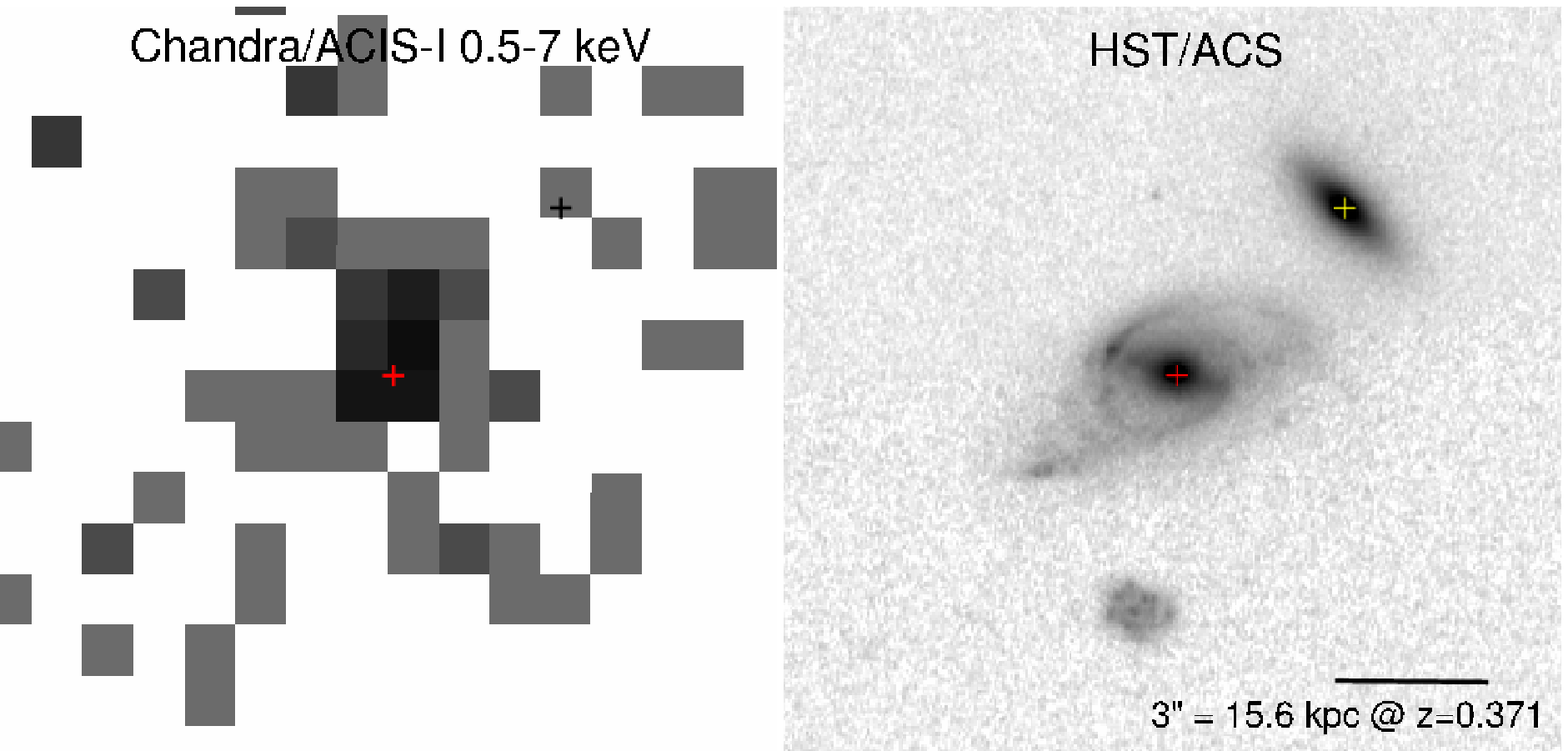}

\caption{AGNs in close kinematic pairs ($left$: $Chandra$; $right$: HST/ACS F814W).  $Top$ An AGN (CID=450) at $z=0.658$ is part of a system of three galaxies within the central potential of an X-ray emitting galaxy group.  The redshifts of the individual galaxies are labelled.  $Middle$ An AGN (CID=1711) is associated with a pair of interacting, spiral galaxies at $z=0.77$.  $Bottom$ A system at $z=0.371$ with a barred spiral hosting an AGN (CID=3083).  In all panels, the optical positions of the zCOSMOS galaxies in pairs are marked with a small cross (red=AGN), and north is up while east is to the left.}
\label{examples}
\end{figure}

We first highlight some examples of AGNs associated with galaxies in close kinematic pairs (Figure~\ref{examples}) that illustrate the diversity in their optical properties.  In the $top$ panel, a system of three galaxies at $z\sim0.659$ is identified, as shown in the HST/ACS F814W image \citep[$right;$][]{ko2007},  to be kinematically linked.  An X-ray source ($left$) is clearly associated with one of the galaxies undergoing an interaction, as is evident from the low-level optical emission linking it to a neighbor (towards the south-east) about 20 kpc away.  In addition to a point-like hard X-ray source, diffuse emission is present and more prominent in the soft energy band as expected (not shown).  In Figure~\ref{examples} ($middle$), two interacting spiral galaxies are separated by $\sim15$ kpc with one of them hosting a luminous AGN while the other is a post-starburst galaxy based on spectral features (i.e., strong Balmer absorption series) evident in a zCOSMOS optical spectrum (not shown).  A third example ($bottom$) clearly shows a barred spiral galaxy with an AGN and a companion 24 kpc away.    

Our objective is to measure the fraction of galaxies that host AGN as a function of whether they are associated with a kinematic pair or not in order to assess whether early encounters between galaxies induce such nuclear activity.  We first perform an analysis based on "observed" galaxies in the zCOSMOS 20k catalog that are actually identified to be in the pair and the field samples.  We note that the field sample includes galaxies, in actual pairs, that have not been identified due to the spectroscopic incompleteness of zCOSMOS thus the observed AGN fraction in non-pair galaxies will likely be overestimated if an excess of AGNs in paired galaxies is realized.  In Section~\ref{intr_results}, we proceed a step further by recovering the intrinsic fraction of AGNs in galaxies in both pairs and the field, including reliable confidence intervals, since we have full knowledge of the selection function of the spectroscopic sample.  This second step further enables us to compute the actual fraction of all AGN that are correlated with the presence of being in a kinematic pair.

For all analysis, we specifically determine whether a zCOSMOS galaxy with $M_*>2.5\times10^{10}$ M$_{\sun}$ is identified with an X-ray detection by $Chandra$.  We allow a galaxy to be associated with a kinematic pair if it has a neighbor satisfying the criteria, given above, on the projected separation and line-of-sight velocity difference.  We further restrict the sample to pairs with a mass ratio less than 10 to 1 which allows for a companion to have a mass below our threshold.  This mass ratio is chosen to isolate a sample for which strong gravitational interactions are capable of destabilizing gas that can subsequently fuel an AGN.  

The observed AGN fraction ($f$; equation{~\ref{fraction}) is determined by summing over the full sample of AGN ($N$) with $N_{eff,i}$ representing the number of effective galaxies in which we could have detected an AGN of a given X-ray luminosity $L^{i}_{\rm X}$ \citep[see][]{si09a,si09b}.  

\begin{equation}
f=\sum_{i=1}^{N}\frac{w_i}{N_{eff, i}}
\label{fraction}
\end{equation}

\noindent This procedure enables us to account for the spatially-varying sensitivity limits of the {\it Chandra} observations (effectively the source detection completeness as a function of X-ray flux shown in Figure 12 of \citealt{puccetti2009}) of the COSMOS field \citep[also see Figure 4 of][]{elvis2009}.  The effective number of galaxies typically falls below the total number of galaxies by $\sim15\%$.  An additional weight ($w_i$) denotes whether an X-ray source was observed as a random ($w_i=1$), or compulsory ($w_i=0.70$) target.  This factor $w_i$ is the ratio of the sampling rate of random targets (0.68) divided by that of the compulsory targets (0.97) within the central zCOSMOS area.  These sampling rates are simply the ratio of observed targets divided by the total number of targets in an input galaxy catalog.  By down weighting the 'compulsory' targets, we effectively remove to first order the bias that some AGNs were targeted at a higher rate as compared to random galaxies.  We also incorporate these weights ($w$) into $N_{eff,i}$ with a minor impact because the majority of galaxies ($\sim97\%$) are random targets.  Due to the fact that the $Chandra$ observations were taken after the zCOSMOS program began, only 30\% of these X-ray sources were compulsory targets, essentially the brighter XMM-$Newton$ sources.  We estimate the associated 1$\sigma$ error \citep[see equation 4 of][]{si09b} based on a gaussian approximation to a beta distribution (\citealt{cameron2010}, Andrae et al. in preparation).

We then compare the fraction of galaxies, in pairs, that host an AGN to those, not in pairs, over the redshift range $0.25<z<1.05$ (Table~\ref{table_results}).  We define an AGN here as an X-ray point source with $\log~L_{0.5-10~{\rm keV}}>42.3$ (units of erg s$^{-1}$), unless otherwise noted; the inclusion of a small number ($13\%$) of more luminous AGNs ($\log~L_X\sim 43.7-44.5$), that may have less accurate stellar mass estimates, has no impact on our results.  As motivated by the analysis given below, we limit the sample of galaxies in pairs to those having a close neighbor that satisfies the criteria that $dr<75$ kpc and $dv<500$ km s$^{-1}$.  We find that over the entire redshift range the AGN fraction for the control sample is $5.0\pm0.4\%$ (not corrected for incompleteness with respect to pair identification; see below), while the pair sample has a higher AGN fraction of $9.8\pm2.2\%$.  While the significance of the difference in the AGN fraction ($\Delta=f_{pairs}-f_{field}$) is $2.2\sigma$ based on simple error propagation, we demonstrate below that the significance is actually higher when considering the sampling rates of the galaxy pairs (Section~\ref{intr_results}).  Even before taking this into consideration, an enhancement of AGN activity is evident for galaxies in pairs having close separations.   

In Figure~\ref{agn_fraction_z}, we show the AGN fraction of galaxies, given this direct simple approach, in kinematic pairs and the control sample as measured in two redshift bins of equivalent width.  We set the minimum X-ray luminosity to roughly the mean sensitivity limit for each redshift bin (see Table~\ref{table_results}).  We find that the fraction of galaxies in pairs hosting an AGN is higher in both redshift intervals, even when implementing a constant luminosity cut (i.e., $\log~L_X>42.6$) over the full redshift range (not shown).  This effectively demonstrates that our result based on the full redshift range is not due to a selection effect incurred by either the host galaxy mass limit, the mass ratio of the pairs, or an offset in the redshift distribution of AGNs and galaxies in pairs.  The increase in the AGN fraction with redshift is a result of the overall evolution of the luminosity function of AGN.

\begin{figure}
\epsscale{1.1}
\plotone{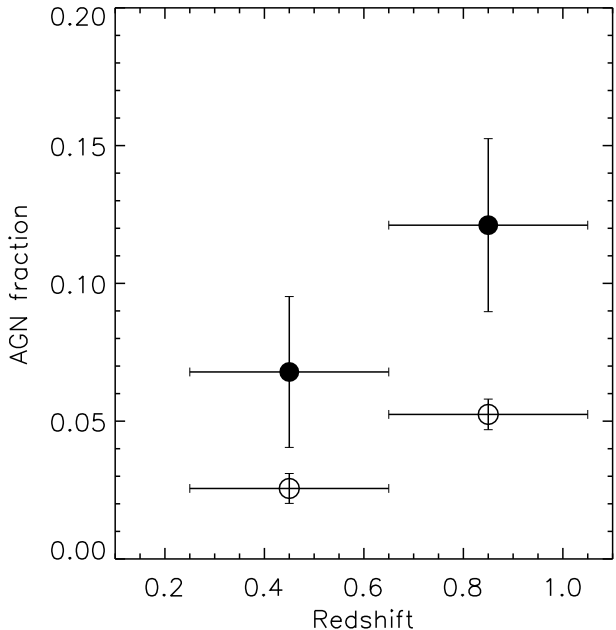}
\caption{Observed AGN fraction of galaxies in close kinematic pairs ($dr < 75$ kpc and $dv < 500$ km s$^{-1}$; filled black circles) as compared to galaxies with no neighbor within a projected separation of 143 kpc and a velocity offset less than 500 km s$^{-1}$ (open circles) for two redshift intervals.  The horizontal bars indicate the redshift range for each value while the vertical bars are the $1\sigma$ error.}
\label{agn_fraction_z}
\end{figure}

\begin{figure*}
\epsscale{1.0}
\plottwo{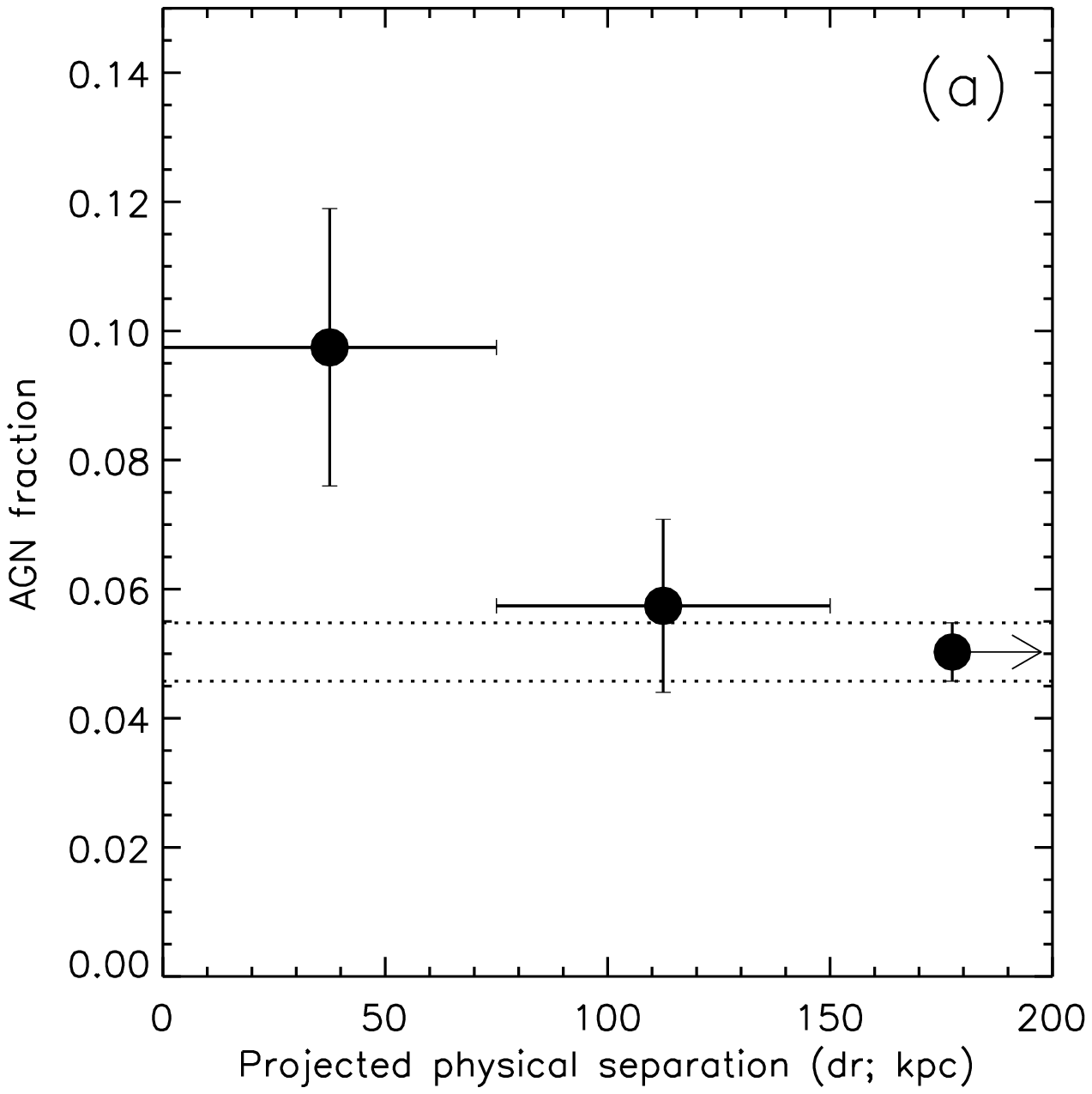}{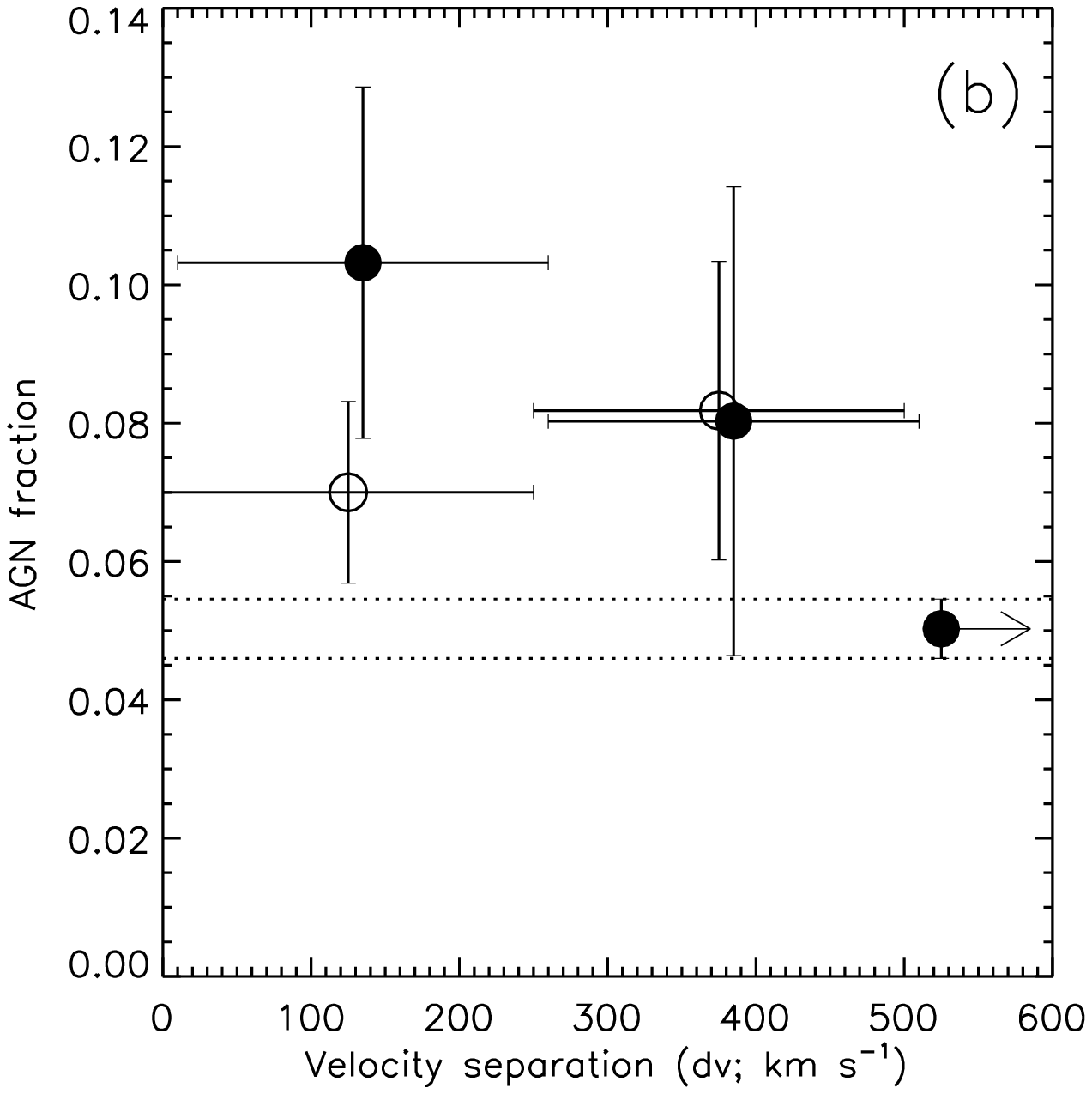}
\epsscale{0.7}
\caption{Observed fraction of galaxies hosting an AGN shown as a function of both the projected physical separation ($a$) and line-of-sight velocity difference ($b$).  The right-most data point in both panels gives the AGN fraction of galaxies not in pairs with the horizontal lines showing the $1\sigma$ error and extended along the abscissa for visual comparison.}
\vspace{0.25cm}
\label{agn_fraction_separation}
\end{figure*}

\begin{deluxetable*}{llllllll}
\tabletypesize{\small}
\tablecaption{Sample statistics-observed kinematic pair analysis\tablenotemark{a}\label{table_results}}
\tablewidth{0pt}
\tablehead{\colhead{Type}&\colhead{Redshift}&dr&dv&$L_{X}$\tablenotemark{b}&\colhead{\# of}&\colhead{\# of}&\colhead{AGN}\\
&&(kpc)&(km s$^{-1}$)&&\colhead{AGN}&\colhead{galaxies}&\colhead{fraction (\%)}
}
\startdata
Pairs&0.25-1.05&$<75$&$<500$&$>2$&19 &223 &$9.75\pm2.15$\\
&0.25-1.05&$<143$&$<500$&$>2$&38 &562  &$7.30\pm1.17$\\
Control&0.25-1.05&$>143$&$>500$&$>2$&116 & 2726 &$5.03\pm0.45$\\
\hline
Pairs&0.25-0.65&$<75$&$<500$&$>2$&7 &102 &$6.75\pm2.56$\\
&0.65-1.05&$<75$&$<500$&$>4$&14 &138 &$11.44\pm2.96$\\
Control&0.25-0.65&$>143$&$>500$&$>2$ & 23 &907 &$2.56\pm0.54$\\
&0.65-1.05&$>143$&$>500$&$>4$&87&1819 &$5.24\pm0.56$\\
\hline
Pairs-random&0.25-1.05&$<75$&$<500$&$>2$&14 &214 &$8.25\pm2.05$\\
&&&&$4-50$&13 &214 &$7.10\pm1.88$\\
Control-random&0.25-1.05&$>143$&$>500$&$>2$&87 & 2671  &$4.28\pm0.43$\\
&&&&$4-50$&63 & 2671  &$2.70\pm0.33$
\enddata
\tablenotetext{a}{Not corrected for incomplete identification of galaxy pairs}
\tablenotetext{b}{units of $10^{42}$ erg s$^{-1}$ (0.5-10 keV)}
\end{deluxetable*}

Our restriction on the projected separation ($dr<75$ kpc) of zCOSMOS pairs, in the above analysis, is due to the fact that we see an enhancement of AGN activity specifically on smaller scales.  To demonstrate this, we measure the AGN fraction of galaxies in pairs while separating the sample into two bins of projected separation ($dr$) and line-of-sight velocity difference ($dv$).  In Figure~\ref{agn_fraction_separation}$a$, we compare the results of this exercise with the AGN fraction of galaxies not in pairs that acts as our control sample.  We find that the AGN fraction of galaxies is highest ($9.8\%$) for galaxies in pairs with $dr \lesssim 75$ kpc and a velocity difference of $dv \lesssim 500$ km s$^{-1}$.  While the AGN fraction is elevated for AGNs in pair galaxies shown as a function of velocity difference (Figure~\ref{agn_fraction_separation}$b$), the level of incidence is not arguably stronger in the bin with the smallest velocity offset.  It does appear that a monotonically increasing AGN fraction (5.0\%, 7.3\%, 9.8\%) with decreasing physical separation is evident when comparing the two dependent samples ($dr<75$ kpc, $dr<143$ kpc) with the control (i.e., field) sample (first three lines in Table~\ref{table_results}).

\subsection{Mass ratios of galaxies in pairs}

We demonstrate that the bulk of AGN activity does occur in zCOSMOS galaxies having a mass ratio of less than 3:1 (i.e., 'major' interaction/merger).  In Table~\ref{table_mass_ratio}, we give the AGN fraction of galaxies with selection in four different mass ratio regimes (i.e., less than the given ratio) along with the size of the AGN sample.  Our AGN selection here includes a limit on the maximum luminosity of the AGN to avoid those that may have inaccurate stellar mass estimates.  We find that 67\% (12/18) of AGNs in galaxy pairs with $dr<75$ kpc are associated with galaxies having a mass ratio within this range.  In Figure~\ref{agn_fraction_mratio},  we plot the AGN fraction as a function of mass ratios of the pairs and include all AGNs with $\log~L_{0.5-8~{\rm keV}}>42.3$.  We remark that these measurements are not independent since they include all galaxies having a mass ratio less than the given value.  There does appear to be a trend of increasing AGN fraction as the masses between the two galaxies are more similar. A larger sample is obviously required to justify such a claim with certainty.  It is reassuring that for each mass ratio we find a consistent result that the AGN fraction increases in kinematic pair sample, as compared to non-pair galaxies (control sample), with the highest levels of incidence occurring on the smaller scales of projected separation.  We note that redshift and mass dependent effects may be non-negligible for pair samples with higher mass ratios ($>10$) since we do not implement a volume or mass limited criteria for both members of a galaxy pair.

\begin{figure}
\epsscale{1.1}
\plotone{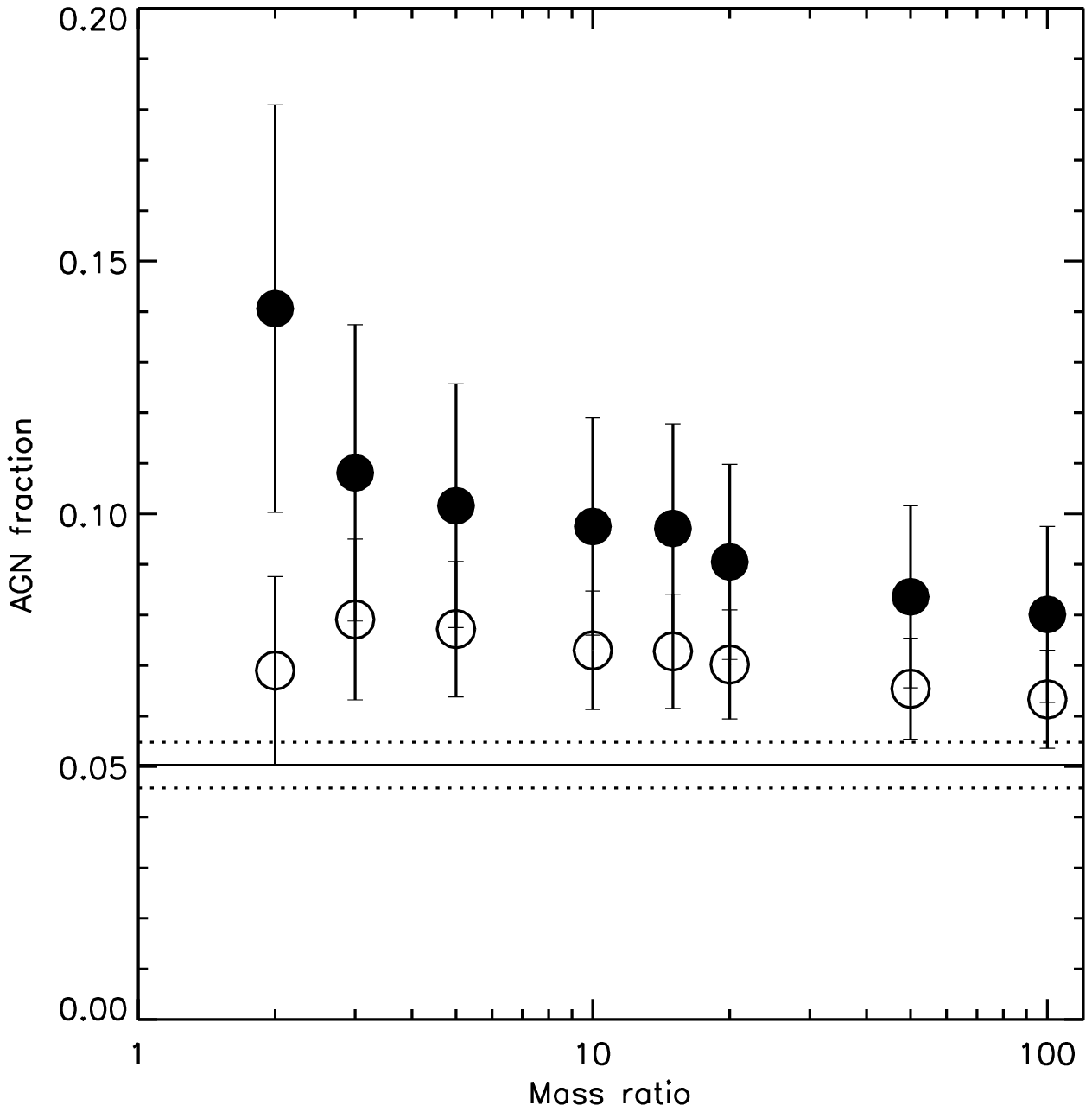}
\caption{Observed AGN fraction of galaxies in close kinematic pairs (filled circles: $dr < 75$ kpc and $dv < 500$ km s$^{-1}$; open circles: $dr < 143$ kpc and $dv < 500$ km s$^{-1}$) as a function of the mass ratio of the two galaxies.  The horizontal bars give the non-pair fraction with $1\sigma$ errors.}
\label{agn_fraction_mratio}
\end{figure}

\begin{deluxetable}{lcc}
\tabletypesize{\small}
\tablecaption{Fraction of galaxies\tablenotemark{a} in pairs of different mass ratios\label{table_mass_ratio} hosting AGN\tablenotemark{b}}
\tablewidth{0pt}
\tablehead{\colhead{Mass ratio\tablenotemark{c}}&\colhead{$dr<75$ kpc}&\colhead{$dr<143$ kpc}\\
&\colhead{\% (\# AGN)}&\colhead{\% (\# AGN)}}
\startdata
15:1&$9.3\pm2.1$ (19)&$6.6\pm1.1$ (35)\\
10:1&$9.8\pm2.2$ (18)&$6.7\pm1.1$ (33)\\
5:1&$11.1\pm2.6$ (17)&$7.4\pm1.3$ (29)\\
3:1&$11.7\pm3.2$ (12)&$7.6\pm1.6$ (21)
\enddata
\tablenotetext{a}{Not corrected for incomplete identification of galaxy pairs}
\tablenotetext{b}{$\log~L_{0.5-10~{\rm keV}} >42.3$}
\tablenotetext{c}{Mass ratio less than the given range.}
\end{deluxetable}

\subsection{Further considerations}

We recognize that a number of selection effects must be considered given their potential impact on the aforementioned results, such as (1) an artificially elevated level of incidence of AGNs in pairs due to differences in the redshift success rate between galaxies and AGNs, or (2) 30\% of the X-ray sources being spectroscopically-observed as compulsory targets.  Any bias in the number of AGNs in our sample can artificially induce a higher fraction of pairs with AGNs if not considered carefully since the probability of having a pair is the product of the sampling rates of the individual galaxies.  To address the first issue,  we measure the success rate of redshift identification with confidence flag $f\geq1.5$ \citep[see][]{lilly09} between randomly-targeted galaxies (89.2\%) and AGNs (91.9\%) in the zCOSMOS sample and find very similar rates.  To address the second issue, we carried out a similar analysis, as done above, to measure the AGN fraction but with a highly conservative selection.  We limited the sample to (1) only galaxies and AGN that were observed as part of the random sample (see Section~\ref{intr_results}).  In addition, we constructed a highly conservative sample that further imposed the following restrictions on the random sample: (2) only AGNs with a higher luminosity cut ($\log~L_X>42.6$) over all redshifts to avoid those falling near the flux limit thus avoiding a large contribution from a few objects, (3) as done above, a limit on the maximum luminosity of the AGN.  Even with these limited samples, we see an enhancement of AGNs on the same scale and with similar confidence (see Table~\ref{table_results}).  We highlight that a stronger enhancement is seen ($2.6\times$) for the case where the X-ray luminosity is limited to the lower end of the distribution ($42.6< \log~L_X<43.7$) that may indicate that such encounters are more likely to result in AGN of moderate luminosity.

It may also be the case that the observed trends are driven by the dependence of AGN activity on stellar mass.  Therefore, we performed a series of K-S tests to determine whether the stellar-mass distribution of the pairs and the control sample differ in any of the bins (i.e., redshift, $dr$, $dv$) implemented in this study.  We find that there is no noticeable difference that can be responsible for the results presented herein.      

\begin{figure*}
\epsscale{1.0}
\plotone{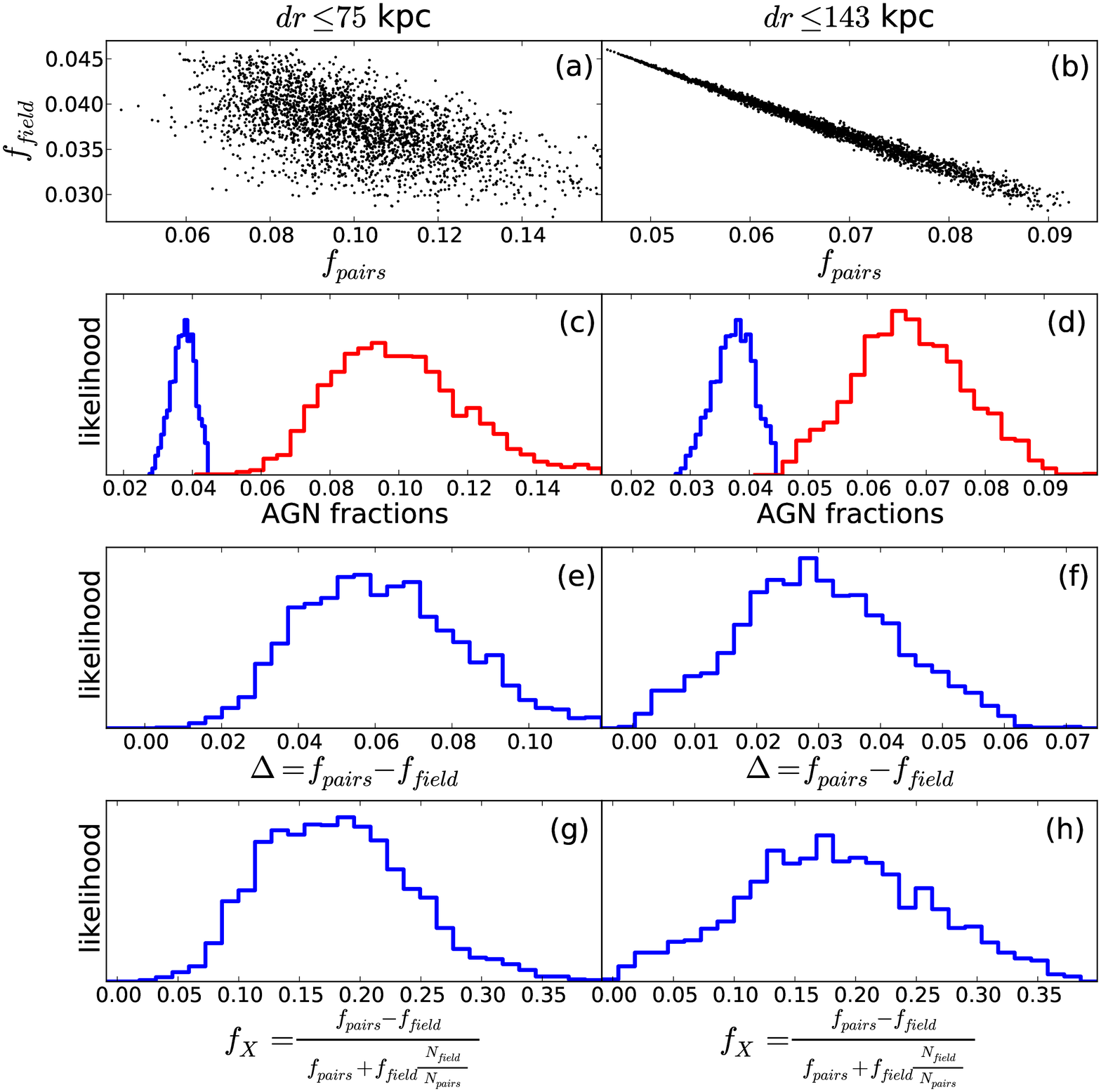}
\caption{Intrinsic fraction of galaxies hosting AGN based on a full Monte Carlo analysis.  The rows are as follows: (1) AGN fraction in the field versus pair sample for the individual realizations, (2) probability distribution for the AGN fraction in the field (blue) and pair sample (red), (3) difference in the AGN fraction between pairs and the field, (4) excess fraction as defined in the text.  The left column refers to pairs with $dr<75$ kpc while the right column shows the equivalent for pairs with $dr<143$ kpc.}
\label{monte_carlo}
\end{figure*}

\section{A Monte-carlo inference of the intrinsic AGN fraction}

\label{intr_results}

As a second step, a complementary analysis, using a Monte Carlo technique as fully outlined in the Appendix, is implemented to recover the intrinsic fraction of AGNs in different subsets of galaxies such as those in galaxy pairs or within a field sample.  This procedure allows us to properly account for (1) the selection effects outlined above, (2) the incompleteness with respect to pair identification (not yet addressed),  and (3) determine confidence intervals for all quantities without an assumption such as Gaussianity.  Accurate knowledge of the fraction of galaxies in a non-pair sample enables us to estimate the impact of mechanisms, not associated with external interactions (i.e., secular processes), responsible for AGN activity and provide an accurate determination of the boost factor of AGNs in pairs relative to the field sample, i.e. the excess of AGN supposedly induced by being in a kinematic pair, after taking out the field "background".

The results of this exercise are shown in Figure~\ref{monte_carlo} and given in Table~\ref{table_intrinsic} (including 68\% and 99\% confidence intervals).  We find a consistent result with the previous analysis for which the AGN fraction of galaxies in pairs with $dr<75$ kpc is 9.7\% (see panel c).  It is now evident that the AGN fraction in the field is significantly reduced to 3.8\%.  Based on these values given in Table~\ref{table_intrinsic} and their confidence intervals, we now find an intrinsic enhancement of AGNs in pairs by a factor of 2.6 with a higher significance ($3.2\sigma$) than previously determined.  Although, an accurate assessment of the significance is non-trivial given the anti-correlation between $f_\textrm{pairs}$ and $f_\textrm{field}$ (Fig.~\ref{monte_carlo}$a,b$).  Also consistent with the previous analysis, we find an increase in the AGN fraction in pairs, spanning the full range of separation ($dr<143$ kpc; Fig.~\ref{monte_carlo}$d$), as compared to the field estimate, although not as strong an enhancement as seen on smaller scales.

\begin{deluxetable*}{llllllll}
\tabletypesize{\small}
\tablecaption{Intrinsic kinematic pair analysis (Monte Carlo approach)\label{table_intrinsic}}
\tablewidth{0pt}
\tablehead{\colhead{Type}&\colhead{Redshift}&dr&dv&$L_{X}$\tablenotemark{a}&\colhead{AGN fraction (\%)}&\colhead{Excess fraction ($f_x$)}\\
&&(kpc)&(km s$^{-1}$)&&(1\%, 16\%, median, 84\%, 99\%)&(1\%, 16\%, median, 84\%, 99\%)}
\startdata
Pairs&0.25-1.05&$<75$&$<500$&$>2$&6.3, 8.0, {\bf 9.7}, 12.0, 15.6&6.1, 11.9, {\bf 17.6}, 24.1, 34.0\\
&0.25-1.05&$<143$&$<500$&$>2$&4.8, 5.8, {\bf 6.7}, 7.7, 8.9&2.2, 10.4, {\bf 17.8}, 26.2, 35.1\\
Field&0.25-1.05&$>143$&$>500$&$>2$&3.0, 3.4, {\bf 3.8}, 4.1, 4.5&---------------------------------
\enddata
\tablenotetext{a}{units of $10^{42}$ erg s$^{-1}$ (0.5-10 keV)}
\end{deluxetable*}

\subsection{Fraction of the AGN population attributed to galaxy interactions}

We can now use our measurement of the intrinsic fraction of galaxies hosting AGN, either associated with kinematic pairs or not, to determine the relative importance of galaxy interactions, as compared to other factors yet to be firmly established.  These other processes are likely to be internal to their host galaxy since we have determined, using the same zCOSMOS galaxies, that AGN activity is also dependent on the stellar mass and star formation rate of its host \citep{si09b} and not strongly dependent on the large-scale environment \citep[$\sim$Mpc;][]{si09a}.  Further, \citet{cisternas2011} demonstrate that the majority (80\%) of AGN hosts in COSMOS are not undergoing major mergers.  We highlight that the enhancement of AGNs in pairs is really a local effect ($\lesssim143$ kpc); the fact that zCOSMOS pairs are more common in regions of heightened large-scale galaxy overdensity \citep{deravel2011,kampczyk2011} does not mean that structures such as galaxy groups are more conducive for AGN activity (e.g., through the accretion of cold gas) beyond increasing the galaxy merger rate.  We specifically find that AGNs are equally likely to reside in optically-selected groups and the 'field' (non-group) population \citep[see figure 11 of][]{si09a}.         

To address this issue, we measure an excess fraction of galaxies in pairs as compared to those in the 'field'.  This ratio ($f_x$) is the number of AGN triggered by interactions ($n_\textrm{x}$) relative the the total AGN population ($n_\textrm{pairs}+n_\textrm{field}$; Equation~\ref{exfrac1}).  The number of AGN in pairs ($n_\textrm{x}$) can be expressed with $n_\textrm{bg}$ (bg=background), the number of AGNs in the pair sample not triggered by interactions.

\begin{equation}
f_x=\frac{n_\textrm{x}}{n_\textrm{pairs}+n_\textrm{field}}
\label{exfrac1}
\end{equation}

\begin{equation}
n_\textrm{pairs}=n_\textrm{x}+n_\textrm{bg}=n_\textrm{x}+f_\textrm{field}\times N_\textrm{pairs}
\label{exfrac2}
\end{equation}

These equations can be combined and arranged to give the following expression.

\begin{equation}
f_x=\frac{f_\textrm{pairs}-f_\textrm{field}}{f_\textrm{pairs}+f_\textrm{field} \times \frac{N_\textrm{field}}{N_\textrm{pairs}}}
\label{exfrac3}
\end{equation}

We can then evaluate this expression through the results of our Monte Carlo approach to measure the intrinsic AGN fraction in the field and in pairs (see Table~\ref{table_intrinsic}).  In the lower two panels of Figure~\ref{monte_carlo}, we plot the likelihood distribution of $f_x$ for two pairs samples ($left$: $dr<75$ kpc; $right$: $dr<143$ kpc).  We find that interactions between galaxies with $dr<143$ kpc account for $17.8_{-7.4}^{+8.4}\%$ of AGN activity in our sample.  

\begin{figure}
\epsscale{1.1}
\plotone{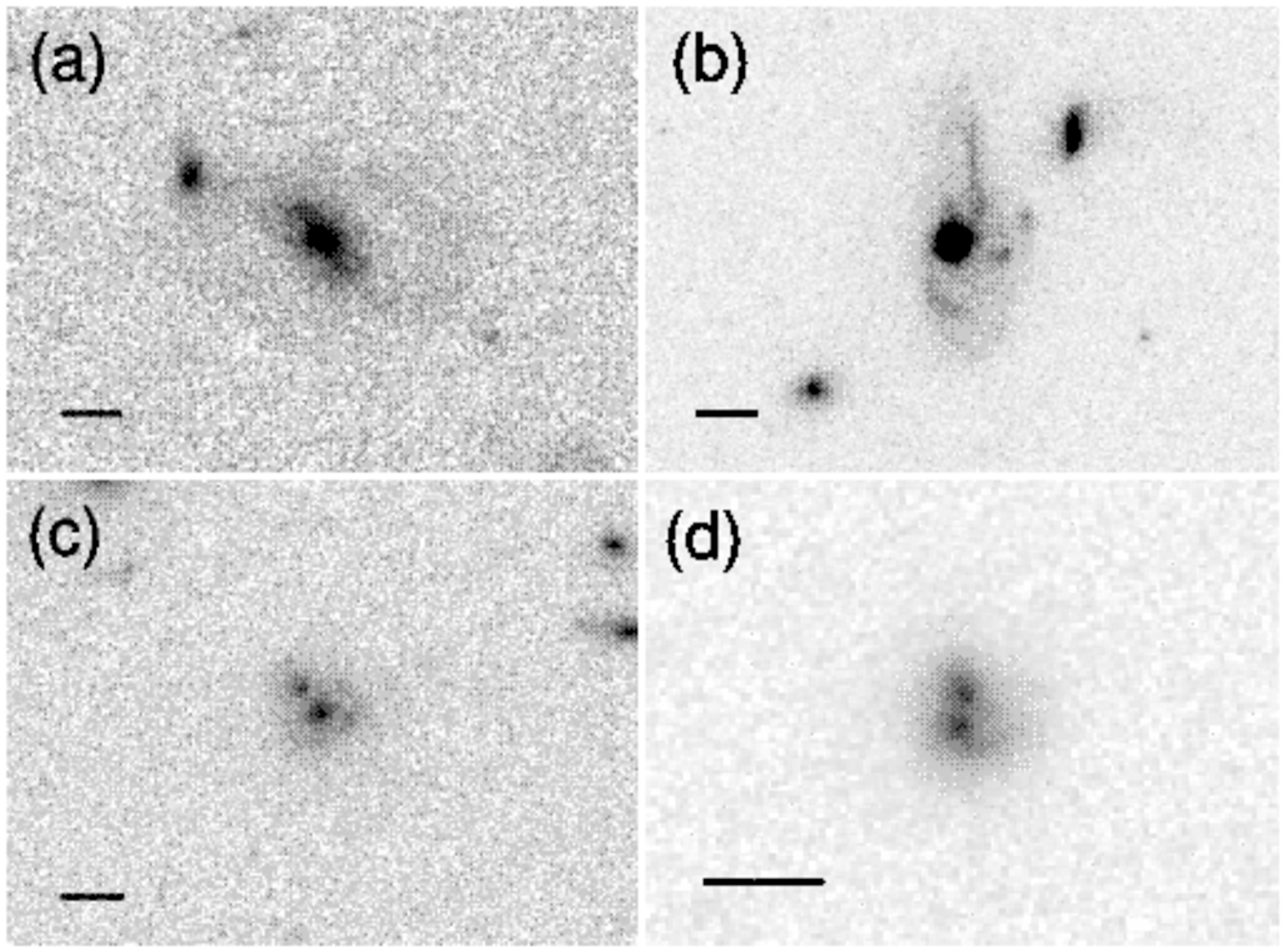}
\caption{HST/ACS images of AGNs that fall within our field sample and identified to be in late stages of a galaxy merger based on their optical morphology of the host galaxy: (a) CID799, (b) CID298, (c) CID401, (d) CID496}
\label{late_mergers}
\end{figure}

\section{Concluding remarks}

We have performed a simple test of the scenario for which galaxy interactions trigger moderate-luminosity AGNs.  To do so, we have utilized a sample of kinematic pairs identified from the zCOSMOS 20k 'bright' catalog and $Chandra$ observations that indicate those harboring AGNs.  The X-ray selection of AGN enables us to include those that may be cloaked in star formation that can hamper optical selection, especially since elevated SFRs are a common property of close galaxy pairs \citep{ellison2008,lin2010,liu2011}.    

Based on the multi-wavelength observations in COSMOS, we have found that galaxies ($0.25<z<1.05$; $M_*>2.5\times10^{10}$ M$_{\sun}$) in kinematic pairs, with physical separations less than 75 kpc and line-of-sight velocity difference less than 500 km s$^{-1}$,  exhibit statistically significant ($3.2\sigma$), intrinsically heightened levels of AGN activity ($9.7_{-1.7}^{+2.3}\% $) relative to galaxies not within these associations ($3.8_{-0.4}^{+0.3}\%$) by a factor of 2.6.  Interestingly, this factor is remarkably close to the boost in star formation seen in zCOSMOS kinematic pairs on small scales \citep{kampczyk2011} thus possibly indicating a common causal connection between galaxy interactions, star formation and AGN activity.

Our observed enhancement of AGNs on these physical scales ($dr<75$ kpc) is in broad agreement with merger-driven models \citep{hopkins2008} of black hole growth.  From these models, the first signs of an obscured AGN should emerge roughly 0.5 Gyr after first passage of two interacting galaxies and of similar physical separation.  Based on a comparison to the Millenium mock catalogs of the COSMOS field, \citet{kampczyk2011} demonstrate that the majority of zCOSMOS kinematic pairs, with such small separations, are likely to merge within a Hubble time.  Keep in mind that our results pertain mainly to moderate-luminosity AGNs \hbox{($L_X\sim10^{43}$ erg s$^{-1}$)}, and in no way are we claiming that kinematic pairs are conducive environments for the more luminous quasars.  Although, it has been shown that SDSS quasars \citep{serber2006} are more strongly clustered on scales less than 100 kpc.

There has been much debate as to whether or not galaxy interactions including major mergers are the single mechanism predominantly driving gas to the nuclear region in galaxies thus inducing accretion onto the central SMBH.  It is now apparent, based on many studies \citep[e.g.,][]{grogin2005,gabor2009,cisternas2011} including the one presented here, that a number of physical processes, yet to be fully understood, are important.  Therefore, we can now attempt to determine the relative contribution of various mechanisms responsible for black hole growth.  Based on our results, we start by estimating the fraction of the moderate-luminosity AGN population that can be attributed solely to galaxy interactions; we find that $\sim18\%$ of such AGNs are the result of interactions.  This leaves open many questions such as the following:  What physical mechanisms are responsible for the remaining $\sim80\%$ of AGNs?  We do recognize that late-stage mergers are not accounted for in our sample.  It is clear that these do exist within the field population (Figure~\ref{late_mergers}) with some exhibiting double optical nuclei, although it is unlikely that they account for a majority the AGN population since \citep{cisternas2011} showed that at $z<1$ the fraction of moderate-luminosity AGN  that are due to the late stages of interaction has an upper limit of $\sim20\%$.  Whether the final coalescence of two massive galaxies results in a luminous quasar remains to be determined and if they emerge at a later times in merger sequence.  Answers to these questions may require further $Chandra$ observations of the full zCOSMOS area or await larger area spectroscopic surveys having a high degree of completeness.    

\acknowledgments

The authors appreciate the useful discussions with Charles Steinhardt, Yen-Ting Lin and Ed Turner.  This work was supported by World Premier International Research Center Initiative (WPI Initiative), MEXT, Japan.  KJ is supported by the Emmy Noether Programme by the German Science Foundation (DFG). RA is funded by the Klaus Tschira Foundation via the Heidelberg Graduate School of  Fundamental Physics (HGSFP). 



Facilities: \facility{CXO(ACIS), VLT:Melipal (VIMOS)}.



\appendix

\section{Monte Carlo simulation of intrinsic AGN fractions}

We construct a method to infer the intrinsic AGN fraction of galaxies either in pairs or the field based on the observed galaxy sample with full consideration of the dominant selection effects.  The analysis is complicated by two special characteristics of our data.  First, galaxies have a detection probability of $w\approx 0.60$, irrespective of whether they host an AGN or not, due to the spectroscopic completeness of the zCOSMOS 'bright' survey \citep[see][]{lilly07,lilly09}.  Here, we define spectroscopic completeness as the ratio of galaxies ($i_{\rm AB}<22.5$) with secure spectroscopic redshifts (confidence flag $\ge1.5$) to all galaxies above this magnitude in our photometric catalog.  As a consequence, galaxies that actually reside in pairs, but were not identified as such (since the partner was not observed), will be present in the field sample.  Second, there is a subsample of $N_\textrm{cs}$ compulsory targets, which are known a priori likely to contain an AGN based on their X-ray or radio emission, although the former is of most concern for our study.  These compulsory targets have detection probability of 1, in contrast to the  other ``random'' targets.   These complications have to be addressed in order to obtain an unbiased estimate of the intrinsic AGN fractions.  While we do account for the differential weighting (compulsory versus random) between targets in Section~\ref{obs_results}, the fact that galaxies in actual pairs can be present in the field sample is not addressed in that analysis.

To fully account for these complexities with the data, we set up a probabilistic model based on a decision tree that accounts for all possible outcomes of an observation. Given an intrinsic total number of galaxies $N_\textrm{tot}^\textrm{int}$ that \textit{could} have been observed down to $i_{\rm AB}=22.5$ within the joint zCOSMOS 'bright' and $Chandra$ survey area, the decision tree considers the following possibilities: \\

\begin{itemize}

\item Is the target random or compulsory? 
\item Has the target been observed?
\item Is the observed target an AGN?
\item Does the observed target reside in a pair? 
\item If the target resides in a pair, has the partner been observed? 

\end{itemize}

\noindent The decision tree has free model parameters that have to be inferred from the data. Among these model parameters are the desired \textit{intrinsic} AGN fractions in the field $f_f$ ($dr>143$ kpc and $dv>500$ km s$^{-1}$), in close pairs $f_{75}$ ($dr<75$ kpc and $dv<500$ km s$^{-1}$) and in loose pairs $f_{143}$ ($75<dr<143$ kpc and $dv<500$ km s$^{-1}$).  Other parameters are the intrinsic total number of galaxies $N_\textrm{tot}^\textrm{int}$, and the probability of a galaxy or AGN to reside in a close pair $p_{75}$ or in loose pairs $p_{143}$. The decision tree then \textit{predicts} the observed quantities. Introducing the number of random targets $N_\textrm{ran} = N_\textrm{tot}^\textrm{int}-N_\textrm{cs}$ and the probability of a target not to reside in a pair but in the field $p_f=1-p_{75}-p_{143}$, the predicted number of observed galaxies in the field is
\begin{equation}\label{eq:NF_pred}
N_f^\textrm{pred} = \left( N_\textrm{cs} + N_\textrm{ran}w\right)\left( p_f + (p_{75}+p_{143})\frac{N_\textrm{ran}}{N_\textrm{tot}^\textrm{int}}(1-w)\right)
\end{equation}
and the predicted number of observed AGN in the field is
\begin{equation}\label{eq:nF_pred}
n_f^\textrm{pred} = N_\textrm{cs}\left( p_f + (p_{75}+p_{143})\frac{N_\textrm{ran}}{N_\textrm{tot}^\textrm{int}}(1-w)\right)
+N_\textrm{ran}w\left( p_f f_f + (p_{75}f_{75}+p_{143}f_{143})\frac{N_\textrm{ran}}{N_\textrm{tot}^\textrm{int}}(1-w)\right) \;\textrm{.}
\end{equation}
Similarly, the predicted number of observed galaxies in close or loose pairs is
\begin{equation}\label{eq:NP_pred}
N_{75/143}^\textrm{pred} = p_{75/143}\left(N_\textrm{cs} + w N_\textrm{ran}\right)\left(\frac{N_\textrm{cs}}{N_\textrm{tot}^\textrm{int}} + w \frac{N_\textrm{ran}}{N_\textrm{tot}^\textrm{int}}\right)
\end{equation}
and the predicted number of observed AGN in close or loose pairs is
\begin{equation}\label{eq:nP_pred}
n_{75/143}^\textrm{pred} = p_{75/143}(N_\textrm{cs}+w\cdot f_{75/143}N_\textrm{ran})\left( N_\textrm{cs} + N_\textrm{ran}w\right) \textrm{.}
\end{equation}
The free parameters of the decision tree, $N_\textrm{tot}^\textrm{int},p_{75},p_{143},f_f, f_{75}$, and $f_{143}$, then have to be estimated in order to match the observed numbers.

In fact, for given observed number of galaxies and AGN in field and pairs, the set of equations given by Eqs.~(\ref{eq:NF_pred}), (\ref{eq:nF_pred}), (\ref{eq:NP_pred}), and~(\ref{eq:nP_pred}) can be solved numerically. These numbers provide an initial guess for the free parameters of the decision tree. While the central-limit theorem suggests that the numerical solution is a good approximation to the most likely values for our parameters, we are also interested in evaluating their uncertainties.  Accurate error estimation is crucial to determine whether an excess of AGN activity in galaxy pairs is realized.  In order to obtain confidence intervals on our free parameters, we need to introduce a likelihood function to our probabilistic model as given by
\begin{displaymath}
\mathcal L(N_\textrm{75}^\textrm{obs},N_\textrm{143}^\textrm{obs},N_\textrm{f}^\textrm{obs},n_ 
\textrm{75}^\textrm{obs},n_ 
\textrm{143}^\textrm{obs},n_\textrm{f}^\textrm{obs}|N_\textrm{tot}^\textrm{int},p_{75},p_{143},f_f, f_{75},f_{143}) =
\end{displaymath}
\begin{displaymath}
=\mathcal L(N_\textrm{75}^\textrm{obs}|N_\textrm{tot}^\textrm{int},p_{75},p_{143},f_f, f_{75},f_{143})\cdot
\mathcal L(n_\textrm{75}^\textrm{obs}|N_\textrm{tot}^\textrm{int},p_{75},p_{143},f_f, f_{75},f_{143})
\end{displaymath}
\begin{displaymath}
\phantom{=}\cdot\mathcal L(N_\textrm{143}^\textrm{obs}|N_\textrm{tot}^\textrm{int},p_{75},p_{143},f_f, f_{75},f_{143})\cdot
\mathcal L(n_\textrm{143}^\textrm{obs}|N_\textrm{tot}^\textrm{int},p_{75},p_{143},f_f, f_{75},f_{143})
\end{displaymath}
\begin{equation}\label{app:eq:factorised_likelihood}
\phantom{=}\cdot\mathcal L(N_f^\textrm{obs}|N_\textrm{tot}^\textrm{int},p_{75},p_{143},f_f, f_{75},f_{143})\cdot
\mathcal L(n_f^\textrm{obs}|N_\textrm{tot}^\textrm{int},p_{75},p_{143},f_f, f_{75},f_{143})
\;\textrm{,}
\end{equation}
which can be factorized because the measurement processes of $N_\textrm{75}^ 
\textrm{obs},N_\textrm{143}^ 
\textrm{obs},N_\textrm{f}^\textrm{obs},n_\textrm{75}^\textrm{obs},n_\textrm{143}^\textrm{obs}$, and  
$n_\textrm{f}^\textrm{obs}$ are statistically independent of each  
other. Of course, the underlying astrophysics imposes relations of  
these observables as described by our probabilistic model. However,  
these relations do \textit{not} lead to correlated measurements. The model predictions of
Eqs.~(\ref{eq:NF_pred},\ref{eq:nF_pred},\ref{eq:NP_pred},\ref{eq:nP_pred}) are always the sums of two binomial variates  
$s=b_1+b_2$, whose likelihood is then given by the convolution of the  
two binomial likelihoods $p(b_i|N_i,f_i)$
\begin{equation}\label{eq:convolution_integral}
\mathcal L(s|N_1,N_2,f_1,f_2) = \int_{\max(0,s-N_2)}^{\min(s,N_1)}p(b| 
N_1,f_1)\,p(s-b|N_2,f_2)\,db \;\textrm{.}
\end{equation}
As $f_1\neq f_2$ in our case, this convolution integral can only be  
solved numerically not analytically, which makes the parameter estimation computationally expensive. For instances, from  
Eq.~(\ref{eq:NP_pred}), we can identify the parameters
\begin{equation}\label{eq:f_pred}
f_1 = p_{75/143}\left(\frac{N_\textrm{cs}}{N_\textrm{tot}^\textrm{int}} + w \frac{N_\textrm{ran}}{N_\textrm{tot}^\textrm{int}}\right)
\quad\textrm{and}\quad
f_2 = p_{75/143}w\left(\frac{N_\textrm{cs}}{N_\textrm{tot}^\textrm{int}} + w \frac{N_\textrm{ran}}{N_\textrm{tot}^\textrm{int}}\right) \;\textrm{.}
\end{equation}
This enables us to evaluate the likelihood function Eq.~(\ref{app:eq:factorised_likelihood}) of the data for given values of the model parameters.

Having identified the likelihood function of the observed data, we can now estimate the model parameters by maximizing this function. Again, since we are particularly interested in assessing differences between the intrinsic AGN fractions, reliable error estimation is of vital importance. Maximum-likelihood parameter estimation is favorable in this case because it provides minimal uncertainties on the results (e.g.,~Barlow 1998). Other routines for parameter estimation are applicable but inevitably produce estimates with larger errors. One option is to draw on the central-limit theorem and approximate the likelihood function by a Gaussian at its maximum. Error estimates can then be derived by Fisher analysis (Heavens 2009). However, we do not want to rely on any such approximations, given the number statistics, especially due to the low AGN fraction of our sample.  Since our probabilistic model contains several free parameters and the evaluation of the convolution integral in Eq.~(\ref{eq:convolution_integral}) is expensive, we employ a Metropolis-Hastings algorithm since a grid search is not an option.  Using such a Markov-Chain Monte-Carlo algorithm has the additional advantage of easily providing marginal estimates $f_{75},f_{143}$, $f_\textrm{f}$, and their differences, which are independent of the other model parameters. This marginalisation is necessary, because otherwise we only obtain conditional error estimates that underestimate the actual errors and would lead us to report overly decisive differences in AGN fractions.


\begin{thebibliography}{}
\bibitem[Bahcall et al.(1997)]{bahcall1997} Bahcall, J. N., Kirhakos, S., Saxe, D. H., Schneider, D. P. 1997, \apj, 479, 642
\bibitem[Barger et al.(2003)]{barger2003} Barger, A. J., Cowie, L.L., Capak, P. et al. 2003, \aj, 126, 632
\bibitem[Bolzonella et al.(2010)]{bo10} Bolzonella, M., Kovac, K., Pozzetti, L. et al. 2009, A\&A, in press, arXiv:0907.0013
\bibitem[Brusa et al.(2010)]{brusa2010} Brusa, M., Civano, F., Comastri, A. et al. 2010, \apj, 716, 348
\bibitem[Bruzual \& Charlot(2003)]{bc03} Bruzual, G., Charlot, S. 2003, \mnras, 344, 1000
\bibitem[Calzetti et al.(2000)]{ca00} Calzetti, D., Armus, L., Bohlin, R. C., Kinney, A. L., Koornneef, J., Storchi-Bergmann, T. 2000, \apj, 533, 682
\bibitem[Cameron(2010)]{cameron2010} Cameron, E. 2010, arXiv:1012.0566
\bibitem[Canalizo \& Stockton(2001)]{canalizo2001} Canalizo, G., Stockton, A. 2001, \apj, 555, 719 
\bibitem[Capak et al.(2007)]{capak07} Capak, P. et al. 2007, \apjs, 2007, 172, 99 
\bibitem[Cisternas et al.(2011)]{cisternas2011} Cisternas, M., Jahnke, K., Inskip, K. J. 2011, \apj, 726, 57
\bibitem[Civano et al.(2010)]{civano2010} Civano, F., Elvis, M., Lanzuisi, G. et al. 2010, \apj, 717, 209
\bibitem[Civano et al.(2011)]{civano2011} Civano, F., Elvis, M. et al. 2011, in preparation
\bibitem[Comerford et al.(2009)]{comerford2009} Comerford, J., Griffith, R. L., Gerke, B.F. et al. 2009, \apj, 702, L82 
\bibitem[de Ravel et al.(2011)]{deravel2011} de Ravel, L., Kampczyk, P., Le Fevre, O., et al. in preparation 
\bibitem[Fu et al.(2010)]{fu2010} Fu, H., Myers, A. D., Djorgovski, S. G., Yan, L. 2010, arXiv:1009.0767 
\bibitem[Gabor et al.(2009)]{gabor2009} Gabor, J., Impey, C., Jahnke et al. 2009, \apj, 691, 705
\bibitem[Green et al.(2010)]{green2010} Green, P.J., Myers, A.D., Barkhouse, W.A. et al. 2010, \apj, 710, 1578 
\bibitem[Grogin et al.(2005)]{grogin2005} Grogin, N. A. et al. 2005, \apj, 627, 97
\bibitem[Hopkins \& Hernquist(2009)]{hopkins2009} Hopkins, P.F., Hernuist, L. 2009, \apj, 694,599
\bibitem[Hopkins et al.(2008)]{hopkins2008} Hopkins, P. F., Cox, T. J., Keres, D., Hernquist, L. 2008, \apjs, 175, 390
\bibitem[Kampczyk et al.(2007)]{kampczyk2007} Kampczyk, P., Lilly, S. J., Carollo, M. C. 2007, \apjs, 172, 329 
\bibitem[Kampczyk et al.(2011)]{kampczyk2011} Kampczyk, P., Lilly, S. J., de Ravel, L. et al. 2011, in preparation 
\bibitem[Kartaltepe et al.(2010)]{kartaltepe2010} Kartaltepe, J. S., Sanders, D. B., Le Floc'h et al. 2010, \apj, 721, 98
\bibitem[Ellison et al.(2008)]{ellison2008} Ellison, S., Patton, D., Simard, L., McConnachie, A. W. 2008, \apj, 135, 1877
\bibitem[Elvis et al.(2009)]{elvis2009} Elvis, M., Civano, F., Vignali, C. et al. \apj, 184, 158
\bibitem[Kewley, Geller \& Barton(2006)]{kewley2006} Kewley, L., Geller, M. J., Barton, E. J. 2006, \aj, 131, 2004
\bibitem[Koekemoer et al.(2007)]{ko2007} Koekemoer, A. et al. 2007, \apj, 172, 196
\bibitem[Li et al.(2008)]{li2008} Li, C., Kauffmann, G., Heckman, T. M., White, S. D., Jing, Y. P. 2008, \mnras, 385, 1915
\bibitem[Lin et al.(2010)]{lin2010} Lin, L., Cooper, M.C., Jian, H-Y. et al. 2010, \apj, 718, 1158
\bibitem[Liu et al.(2011)]{liu2011} Liu, X., Shen, Y., Strauss, M. 2011, arXiv:1104.0951 
\bibitem[Lilly et al.(2007)]{lilly07} Lilly, S. J. et al. 2007, \apjs,  172, 70
\bibitem[Lilly et al.(2009)]{lilly09} Lilly, S. J. et al. 2009, \apjs, 184, 218
\bibitem[Lynden-Bell \& Rees(1971)]{lynden-bell1971} Lynden-Bell, D., Rees, M. 1971, \mnras, 152, 461
\bibitem[Mainieri et al.(2007)]{mainieri2007} Mainieri, V., Hasinger, G., Cappelluti, N. et al. 2007, \apjs, 172, 368
\bibitem[Mihos \& Hernquist(1996)]{mihos1996} Mihos, C. J., Hernquist, L. 1996, \apj, 464, 641
\bibitem[Pozzetti et al.(2010)]{po10} Pozetti, L., Bolzonella, M., Zucca, E. et al. 2009, A\&A, in press, arXiv:0907.5416
\bibitem[Puccetti et al.(2009)]{puccetti2009} Puccetti, S., Vignali, C., Cappelluti, N. et al. 2009, \apjs, 185, 586
\bibitem[Salpeter(1964)]{salpeter1964} Salpeter, E. E. 1964, \apj, 140, 796
\bibitem[Sanders \& Mirabel(1996)]{sanders1996} Sanders, D. B. \& Mirabel, I. F. (1996). Luminous Infrared Galaxies, \textit{ARA\&A} \textbf{34}, 749
\bibitem[Scoville et al.(1989)]{scoville1989} Scoville, N., Sanders, D. B., Sargent, A. I. 1989, \apj, 345,25
\bibitem[Scoville et al.(2007)]{sco07} Scoville, N., Aussel, H., Brusa, M. et al. 2007, \apjs, 172, 1
\bibitem[Szokoly et al.(2004)]{szokoly2004} Szokoly, G., Bergeron, J., Hasinger, G. 2004, \apjs, 155, 271
\bibitem[Serber et al.(2006)]{serber2006} Serber, W., Bahcall, N., M\'{e}nard, B., Richards, G.. 2006, \apj, 643, 68 
\bibitem[Schawinski et al.(2010)]{schawinski2010} Schawinski, K., Urry, M. C., Virani, S. et al. 2010. \apj, 711, 284
\bibitem[Silverman et al.(2008)]{silverman2008} Silverman, J. D., Mainieri, V., Lehmer, B. D. et al. 2008, \apj, 675, 1025
\bibitem[Silverman et al.(2009a)]{si09a} Silverman, J.D., Kovac, K., Knobel, C. Lilly, S. J., et al. 2009a, \apj, 695, 171  
\bibitem[Silverman et al.(2009b)]{si09b} Silverman, J. D., Lamareille, F., Maier, C. et al. 2009b, \apj, 696, 396
\bibitem[Silverman et al.(2010)] {silverman2010}Silverman, J. D., Mainieri, V., Salvato, M. et al. 2010, \apjs, 191, 124
\bibitem[Shen et al.(2010)]{shen2010} Shen, Y., Liu, X., Greene, J., Strauss, M. 2010, in press, arXiv:1011.5246
\bibitem[Xue et al.(2010)]{xue2010} Xue, Y., Brandt, W. N., Luo, B. et al.  2010, \apj, 720, 368
\end{thebibliography}
\end{document}